\documentclass[a4paper,fleqn,review]{cas-sc}

\usepackage[authoryear]{natbib}
\usepackage{hyperref}
\usepackage{subcaption}
\usepackage{makecell}
\usepackage{adjustbox}
\usepackage{diagbox}
\usepackage{gensymb}
\usepackage{amssymb}
\usepackage{xcolor}
\hypersetup{
citebordercolor=white,
linkbordercolor=white,
filebordercolor=white,
urlbordercolor=white}
\usepackage{amsmath}
\usepackage{amsfonts}
\usepackage{mathtools}
\usepackage{graphicx}
\usepackage{lineno}
\newcommand{\mm}[1]{\mathrm{#1}} 

\newcommand{\Xs}{$X_{\mm{S},0}$ }

\newcommand{\rcr}{$\frac{r_\mm{c}}{r}$ }

\newcommand{\Xsp}{$X_{\mm{S},0}$}

\newcommand{\rcrp}{$\frac{r_\mm{c}}{r}$}

\begin{document}

\let\WriteBookmarks\relax
\def\floatpagepagefraction{1}
\def\textpagefraction{.001}

\shorttitle{}

\shortauthors{H.R. Sanderson et~al.}

\title[mode = title]{Long-lived thermal dynamo generation on differentiated, impact-disrupted planetesimals} 

%
\author[1,2]{Hannah R. Sanderson}[orcid = 0000-0001-5842-6985]

\cormark[1]

\fnmark[1]

\ead{hannah.sanderson@geo.uio.no}

\credit{Conceptualization, Methodology, Software, Writing - Original Draft, Writing - Reviewing \& Editing}

\affiliation[1]{organization={Centre for Planetary Habitability},
    addressline={Department of Geosciences, University of Oslo}, 
    city={Oslo},
    country={Norway}}
\affiliation[2]{organization={Department of Earth Sciences},
    addressline={University of Oxford, South Parks Road}, 
    city={Oxford},
    postcode={OX1 3AN}, 
    country={UK}}

\author[2]{James F.J. Bryson}[orcid = 0000-0002-5675-8545]
\credit{Conceptualization, Writing - Review \& Editing, Supervision}

\author[2]{Claire I.O. Nichols}[orcid = 0000-0003-2947-5694]

\credit{Conceptualization, Writing - Review \& Editing, Supervision}

\cortext[cor1]{Corresponding author}

\begin{abstract} 
Meteorites containing both metal and silicates indicate that some planetesimals were partially differentiated and/or processes mixed planetesimal cores and mantles post-differentiation. Time-resolved paleomagnetic records in these meteorite groups can shed light on the differentiation and mixing histories of such bodies.
Previous studies measured paleomagnetic remanences in two such meteorite groups, the IIE irons and the Main Group pallasites, and used thermal evolution and dynamo generation modelling to recover parent body properties. However, these studies assumed that these meteorites only recorded magnetic fields generated by core solidification; an assumption recently shown to be invalid.
We use a refined planetesimal thermal evolution and dynamo generation model that considers thermal and compositional drivers of dynamo generation simultaneously to re-evaluate the parent body properties of the IIE irons and Main Group pallasites and constrain the formation mechanisms of these meteorites. 
We find that none of the remanences require core solidification, but later-formed remanences are more likely to record dynamos driven by core solidification. The Main Group pallasite and IIE iron parent bodies likely had radii of $\sim$400\,km with core radius fractions of $\sim$0.5 and $\sim$0.7, respectively. Impacts shaped both parent bodies: the Main Group pallasites formed too far from the core-mantle-boundary to result from ferromagmatism and instead formed by impacts, and the IIE iron’s core radius fraction suggests the parent body experienced mantle-stripping collisions.
Overall, combining meteorite paleomagnetism with thermal evolution and dynamo generation models provides insights into the long-term evolution of differentiated planetesimals, their interior structures, and metal-silicate mixing on these bodies.
\end{abstract}


\begin{highlights}
\item The IIE irons and Main Group pallasites could record fields from a thermal dynamo
\item The Main Group pallasite and IIE iron parent bodies were large ($\sim$400\,km radius)
\item Mantle-stripping impacts increased the IIE iron parent body’s core radius fraction
\item The Main Group pallasites formed far from the CMB by impacts not ferromagmatism
\end{highlights}


\maketitle

\section{Introduction}
The earliest accreting planetesimals \citep[$<$2\,Ma after CAI formation;][]{spitzer_nucleosynthetic_2021,bryson_collective_2026} contained sufficient $\rm^{26}Al$ to melt and differentiate, forming metallic cores and rocky mantles. However, some meteorites consist of mixtures of metal and silicates on centimetre lengthscales suggesting some planetesimals experienced either incomplete differentiation or later mixing between core and mantle material \citep{krot_11_2014,ruzicka_silicate-bearing_2014}. One example is iron meteorites with chondritic or achondritic silicate inclusions, such as the IAB and IIE irons \citep{ruzicka_silicate-bearing_2014}. Another example is the pallasites, which consist of achondritic silicates surrounded by an Fe-Ni matrix \citep{wasson_main-group_2003}. Pallasites have been proposed to form at the core-mantle boundary \citep[CMB, e.g.,][]{scott_formation_1977,scott_pallasitesmetal_1977,wasson_main-group_2003}, or through impacts between differentiated planetesimals \citep[e.g.,][]{yang_main-group_2010,tarduno_evidence_2012,walte_two-stage_2020,kruijer_tungsten_2022}, or through intrusion of metallic material from the core into the mantle \citep{johnson_ferrovolcanism_2020}. 

Meteorite paleomagnetism can help to reveal the properties of these meteorite parent bodies, and shed light on their formation in several ways. Firstly, a non-zero paleomagnetic remanence that dates from $>$5\, Ma after CAI formation argues strongly that the parent body had a core \citep[e.g.,][]{carporzen_magnetic_2011,elkins-tanton_chondrites_2011,weiss_differentiated_2013,maurel_meteorite_2019}. Secondly, a non-zero remanance strong enough to be generated by a dynamo also indicates a meteorite cooled sufficiently to reach its blocking temperature (i.e., record a remanence) while a magnetic field was being generated. This precludes the formation of a meteorite at the CMB, where temperatures would have been far above the blocking temperature of the magnetic minerals in meteorites when the dynamo was being generated \citep{tarduno_evidence_2012}. Thirdly, a time-resolved paleomagnetic record combined with meteorite cooling rates can trace the thermal evolution of a planetesimal's interior. This information has previously been used to constrain the size of the parent body, its core size, and the timing of core crystallisation \citep[e.g.,][]{bryson_long-lived_2015,maurel_long-lived_2021,nichols_time-resolved_2021}. Collecting these constraints from multiple parent bodies can reveal the structure and evolution of differentiated bodies throughout the early Solar System.  

Two meteorite groups that contain both metal and silicates and have time-resolved paleomagnetic records are: the IIE irons \citep{maurel_meteorite_2020,maurel_long-lived_2021}; and the Main Group pallasites \citep{tarduno_evidence_2012,bryson_long-lived_2015,nichols_pallasite_2016,nichols_time-resolved_2021}. Paleomagnetic measurements have been performed on two carriers within these meteorites: single crystal paleointensity analysis on olivine crystals containing taenite inclusions in the Main Group pallasites \citep{tarduno_evidence_2012,nichols_time-resolved_2021}; and X-ray photoemission electron microscopy (XPEEM) on cloudy zones within meteorite metal in both the Main Group pallasites and IIE irons \citep{bryson_long-lived_2015,nichols_pallasite_2016,maurel_meteorite_2020,nichols_time-resolved_2021,maurel_long-lived_2021,mansbach_resolving_2026}. Cloudy zones are Fe-Ni microstructures that form within meteorite metal by spinodal decomposition below 623\,K in slow cooled meteorites \citep[$\rm<10,000\,KMa^{-1}$;][]{maurel_meteorite_2019}. As the cloudy zone cools through 593\,K (its ordering temperature), it records a chemical transformation remanent magnetisation (CTRM) that is stable for billions of years \citep{einsle_nanomagnetic_2018}. A time-resolved paleomagnetic record can be obtained by measuring several meteorites with different cooling rates from a single parent body. The slower the cooling rate, the deeper within the body the meteorite originated, and so the later it cooled through its ordering temperature and acquired a remanence. Two cooling rates can be obtained from the microstructures in metal alloys in each meteorite: one at 875--975\,K when kamacite nucleated \citep{yang_main-group_2010} and another at 623\,K when the cloudy zone formed \citep{yang_new_1997,goldstein_thermal_2009,maurel_meteorite_2019}.

The time-resolved records of the Main Group pallasites and the IIE irons contain five and three meteorites, respectively. For Main Group pallasites the first two of these meteorites to cool through the cloudy zone ordering temperature (Marjalahti and Brenham) carry null magnetic remanences consistent with the absence of a dynamo field \citep{nichols_pallasite_2016,maurel_meteorite_2019}. Springwater \citep{nichols_time-resolved_2021}, Imilac, and Esquel \citep{tarduno_evidence_2012,bryson_long-lived_2015} acquired their remanences later and all record non-zero paleointensities. Planetesimal dynamo generation models predict that this behaviour results from two epochs of dynamo generation with a gap in between \citep{sanderson_unlocking_2025,sanderson_early_2024}, suggesting that the dynamo restarted between Brenham and Springwater acquiring their remanences. In contrast, all three IIE irons that have been paleomagnetically measured (Colomera, Techado and Miles) record non-zero paleointensities \citep{maurel_meteorite_2020,maurel_long-lived_2021}. 

Planetesimal thermal evolution and dynamo generation models have been used to estimate the formation depths of these meteorites by finding depths that match metallographic cooling rates  \citep{tarduno_evidence_2012,bryson_long-lived_2015,nichols_time-resolved_2021,maurel_meteorite_2020,murphy_quinlan_conductive_2021,maurel_long-lived_2021}. For the Main Group pallasites, whose remanences are not radiometrically dated, the timing of remanence acquisition has been inferred by calculating when these depths cooled through their ordering temperature. Parent body radii and core radius fractions are constrained to values that produce a dynamo at the time that non-zero remanences are recorded. The most recent study of the Main Group pallasites, and the first to include all five meteorites, concluded that the Main Group pallasite parent body had a large core and thin mantle (core radius $\gtrsim70$\% of the total parent body radius) and was 180--360\,km in radius \citep{nichols_time-resolved_2021}. Similar modelling for the IIE irons suggested that the parent body had a radius $\geq$220\,km and was partially differentiated with a core radius fraction $\geq19\%$ \citep{maurel_meteorite_2020,maurel_long-lived_2021}.

These previous models for the Main Group pallasites and IIE irons assumed that these meteorites could only have recorded magnetic fields generated by compositional convection during core solidification and did not consider thermal dynamo generation prior to core solidification. Additionally, many models used the duration of eutectic core solidification as a proxy for dynamo generation \citep[][]{tarduno_evidence_2012,maurel_meteorite_2020,maurel_long-lived_2021,nichols_time-resolved_2021} rather than calculating whether compositional convection was sufficiently vigorous for dynamo generation \citep[using the magnetic Reynolds number, e.g.,][]{nimmo_energetics_2009}. Models that included non-eutectic sulfur contents \citep{bryson_long-lived_2015,nichols_time-resolved_2021} only considered outward core solidification, whereas planetesimal cores likely solidify inwards \citep{williams_bottom-up_2009,dodds_direction_2025}. 

\citet{sanderson_early_2024} used a refined planetesimal thermal evolution and dynamo generation model \citep{sanderson_unlocking_2025} that can simultaneously consider thermal and compositional drivers of dynamo generation, to demonstrate that later remanences do not require core solidification, regardless of whether there is a gap in dynamo generation, and instead could have been generated by thermal convection. This model assumes inward core solidification and can consider a range of core sulfur contents. \citet{sanderson_early_2024} showed that variations in mantle viscosity parameters and planetesimal size can lead to longer dynamo durations than previously considered. Here, we use this refined model to re-evaluate the depths and timings of remanence acquisition and the properties of the Main Group pallasite and IIE iron meteorite parent bodies. We simultaneously vary multiple planetesimal parameters, such as planetesimal viscosity and radius, to account for parameter uncertainties when constraining parent body properties. Altogether, this study sheds light on the size and core radius fraction of differentiated planetesimals and the formation mechanisms for stony-iron and silicate-bearing iron meteorites.

\section{Methods}\label{met}
We calculated possible parent body thermal and magnetic histories using the thermal evolution and dynamo generation model presented by \citet{sanderson_unlocking_2025}. We also included the amended mantle viscosity and mantle solidus from \citet{sanderson_dynamo_2026} that account for non-zero mantle water contents in nominally anhydrous minerals. We explored 90,000 planetesimal parameter combinations (Section \ref{met:params}) and compared the simulated planetesimal history for each parameter combination with measured cooling rates and paleomagnetic remanences of meteorites the Main Group pallasite and IIE iron parent bodies to assess whether it was consistent with the observations (see Sections \ref{met:iie} and \ref{met:pallasite}). We collated planetesimal parameter combinations that produced histories consistent with the observations to determine the distribution of possible parent body parameters and meteorite depths within each parent body, and possible remanence acquisition times for the Main Group pallasites, which do not have radiometric age constraints.

\subsection{IIE iron constraints}\label{met:iie}
Three sets of measurements on the IIE iron meteorites provide constraints on the thermal and dynamo history of their parent body (Table \ref{tab:comb-con}). Firstly, the thermodynamics of cloudy zone formation provides an estimate of each meteorite's cooling rate at 623\,K \citep{maurel_meteorite_2019,maurel_meteorite_2020,maurel_long-lived_2021}. Secondly, the closure temperature of $\rm ^{40}Ar/^{39}Ar$ system in 0.1--1\,mm feldspars overlaps with the cloudy zone ordering temperature \citep[$603\pm70$\,K;][]{bogard_chronology_2000,cassata_argon_2011}, so can date remanence acquisition. Thirdly, all three meteorites record a paleomagnetic remanence, indicating the planetesimal generated a dynamo field at the times that the cloudy zone in each meteorite cooled through 593\,K \citep{maurel_meteorite_2020,maurel_long-lived_2021}. For a model thermal and magnetic history to be a possible fit to the IIE parent body, depths that cool through 593\,K at each meteorite's $\rm ^{40}Ar/^{39}Ar$ age must have cooling rates at 623\,K consistent with the observed values. Also, for each meteorite, the model paleointensity must lie within the recovered paleointensity range at each meteorite's $\rm ^{40}Ar/^{39}Ar$ age. In this study, we compare normalised model and observed paleointensities rather than absolute values to mitigate the uncertainties from magnetostatic interactions and induced fields around cloudy zones \citep{maurel_meteorite_2020,maurel_long-lived_2021,mansbach_resolving_2026} as well as uncertainties on dynamo scaling laws \citep{sanderson_unlocking_2025}. Observed paleointensities were normalised to the strongest recovered paleointensity value and model magnetic field strengths were normalised to the strongest model value within the $\rm ^{40}Ar/^{39}Ar$ age ranges for the timing of remanence acquisition (for more information see Section S1.3).

\subsection{Main Group pallasite constraints}\label{met:pallasite}
For the Main Group pallasites the cooling rate at the kamacite nucleation temperature \citep[875--975\,K, 925\,K applied as an average;][]{yang_main-group_2010} and the cooling rate at the onset of cloudy zone formation \citep[623\,K;][]{maurel_meteorite_2019} provide constraints on the parent body's thermal history (Table \ref{tab:comb-con}) and possible meteorite formation depths must match the observed cooling rates at both these temperatures. The dynamo must be active/inactive as required by the paleomagnetic observations for each meteorite when its formation depth cools through the cloudy zone ordering temperature (593\,K). Unlike for IIE irons, we choose not to utilise relative paleointensities as a constraint, because of large uncertainties on the paleointensities of Imilac and Esquel. In the methodology applied to these meteorites, their paleomagnetic vectors were only partially constrained so paleointensity or direction could not be reliably recovered \citep[][for further discussion see Supplementary Materials]{bryson_paleomagnetic_2019,nichols_time-resolved_2021}.

\begin{table}[htbp]
\resizebox{\columnwidth}{!}{
    \centering
    \begin{tabular}{|c|c|c|c|c|c|c|c|c|}\hline
   & \multicolumn{3}{|c|}{IIE irons} & \multicolumn{5}{|c|}{Main Group pallasites}\\\hline
       & Techado & Colomera & Miles  & Marjalahti & Brenham & Springwater & Imilac & Esquel \\\hline
       \makecell{$\rm ^{40}Ar/^{39}Ar$ age$^{1}$\\/Ma after CAI formation} & $78\pm13$ & $97\pm10$& $159\pm9$ & -- & -- & -- & -- & --\\
        \makecell{Cooling rate at 925\,K$^{2}$  \\/$\rm KMa^{-1}$} &-- &-- & -- &$7.6\pm0.6$ & $6.2\pm0.9$ & $5.1\pm0.7$ & $4.3\pm0.3$ & $3.3\pm0.6$ \\
        \makecell{Cooling rate at 623\,K$^{3,4,5}$ \\/$\rm KMa^{-1}$} & $4.6\pm1.9$ & $2.5\pm1.4$ & $3.8\pm2.6$ & $2.9\pm1.5$ & $2.5\pm1.4$ & $1.7\pm1.2$ & $1.2\pm0.7$ & $0.9\pm0.5$ \\
        Paleointensity$^{3,4,5,6}$ /$\mu \rm T$ & $36\pm11$ & $15\pm5$ & $33\pm9$ & 0& 0& $22\pm8$ & 4.8 -- 95 & 3.3 -- 79\\
      Normalised paleointensity & $1.0\pm0.4$ & $0.42\pm0.18$ & $0.92\pm0.38$ & --& --& --&-- & -- \\
        Dynamo behaviour$^{3}$ & On & On & On & Off & Off & On & On & On \\\hline
    \end{tabular}}
    \caption{Experimental constraints on parent body histories for the IIE irons and Main Group pallasites. $^1$\citet{bogard_chronology_2000}, $^3$\citet{maurel_meteorite_2020},$^4$\citet{maurel_long-lived_2021} $^2$\citet{yang_main-group_2010}, $^5$\citet{maurel_meteorite_2019},$^6$\citet{nichols_time-resolved_2021}. -- indicate values were not available for that meteorite. Main Group pallasite dynamo behaviour is limited to whether the dynamo is on or off due to the uncertainty in the paleointensity measurements of Imilac and Esquel \citep{nichols_time-resolved_2021}. The mean paleointensities for Brenham and Marjalahti are less than the zero-field threshold so are listed as zero \citep{maurel_meteorite_2019}.}
    \label{tab:comb-con}
\end{table}

\subsection{Parameter variation}\label{met:params}
We varied planetesimal radius, core radius fraction, mantle viscosity \citep[reference viscosity, critical melt fraction, Arrhenius slope, melt weakening exponent; for more details see][]{sanderson_early_2024}, initial core sulfur content, and mantle water content in nominally anhydrous minerals (NAMs). We ran models for all possible combinations of parameter values in Table \ref{tab:params} (90,000 runs). To enable this to be computationally feasible, we used 3--5 parameter values within each range. We held radiogenic $\rm^{60}Fe$ abundance and differentiation time constant, because they only affect the first 5\,Ma of dynamo generation \citep{sanderson_early_2024,sanderson_dynamo_2026}, which is well before the IIE or Main Group pallasites could have recorded a remanence. We held liquid viscosity constant, because it has no effect on the timing of dynamo generation \citep{sanderson_early_2024}. Small planetesimals ($\leq100$\,km radius) were not considered, because they are too small to produce the long-lived dynamo generation required \citep{sanderson_early_2024}. The range of reference viscosities, critical melt fractions, and melt-weakening exponents was taken from \citet{sanderson_early_2024}. The upper limit on the Arrhenius slope was slightly raised compared to \citet[][0.035 to 0.05]{sanderson_early_2024} to include higher values due to water content in NAMs \citep{sanderson_dynamo_2026}. The range of water contents in NAMs was also taken from \citet{sanderson_dynamo_2026}. 

The core must be molten at the beginning of a model run, so the initial core sulfur content is restricted to those which have liquidus temperatures below the differentiation temperature \citep[for further discussion see][]{sanderson_unlocking_2025}. The differentiation temperature depends on the water content in NAMs, the critical melt fraction, and planetesimal size. Therefore, the range of initial core sulfur contents allowed for all possible combinations of other parameters is narrow (28.7--32\,wt\%). As a result, we performed an additional series of runs with a wider range of initial core sulfur contents where all other parameters were held constant  at their modal values from the initial parameter exploration (Figures \ref{fig:iie-in-hist}, \ref{fig:pal-in-hist} and Table \ref{tab:params}). The range of initial core sulfur contents for the Main Group pallasites (24--33\,wt\%) is narrower than for the IIE irons (13-33\,wt\%), because of the lower modal critical melt fraction (Figures \ref{fig:iie-in-hist}, \ref{fig:pal-in-hist}) and resulting lower differentiation temperature for the Main Group pallasites. 

\begin{table}[htb]
    \centering
    \resizebox{\columnwidth}{!}{
    \begin{tabular}{|c|c|c|c|c|}\hline
        Parameter &  Symbol & Parameter values & IIE mode & \makecell{Main Group\\Pallasite mode} \\\hline
        Radius /km & $r$ & 200, 300, 400, 500 & 400 & 400 \\
        Core radius fraction & \rcr &  0.1, 0.3, 0.5, 0.7, 0.9  & 0.7 & 0.5 \\
        Initial core sulfur content /wt\% & \Xs & 28.7, 30.35, 32 & 32 & 32 \\
        Water content in NAMs /wt\% & $X_{\mm{w}}$ & 0, 0.2, 0.5, 0.7& 0 & 0 \\
        Critical melt fraction & $\phi_C$ & 0.3, 0.4, 0.5  & 0.4 & 0.3 \\
        Reference viscosity /Pas & $\eta_0$ & $10^{15}$, $10^{17}$ , $10^{19}$ , $10^{21}$, $10^{23}$ &$10^{23}$ & $10^{23}$ \\
        Arrhenius slope /$\rm K^{-1}$ & $\beta$ &  0.01, 0.02, 0.03, 0.04, 0.05 & 0.03 & 0.05 \\
        Melt weakening exponent & $\alpha_{\mm{n}}$ & 25, 30, 35, 40, 45 & 35 & 45\\\hline
    \end{tabular}}
    \caption{Values for the parameter-space explorations. The parameter ranges are from \citet{sanderson_early_2024} and \citet{sanderson_dynamo_2026} or are described in the text.  Liquid viscosity, $\rm^{60}Fe/^{56}Fe$, and differentiation time were held constant at 100\,Pas, $10^{-8}$ and 1\,Ma after CAI formation, respectively. NAMs = nominally anhydrous minerals.}
    \label{tab:params}
\end{table}
\clearpage
\section{Results}\label{res}
\subsection{Parent body properties}\label{res:pb-props}
Of all the simulations we conducted, 1.3\% were consistent with the IIE parent body history and 0.96\% were consistent with the Main Group pallasite parent body history, because the constraint on the IIE irons' magnetic history is more relaxed. The IIE irons record a continuous period of dynamo generation from $\sim$60--170\,Ma after CAI formation, which could have occurred during the first or second epoch of dynamo generation \citep{sanderson_early_2024}. In contrast, the Main Group pallasites record a period without a dynamo followed by a period with a dynamo, so remanences must have been acquired after the gap in dynamo generation predicted by the model.

\begin{figure}[htbp]
    \centering
    \includegraphics[width=0.8\textwidth]{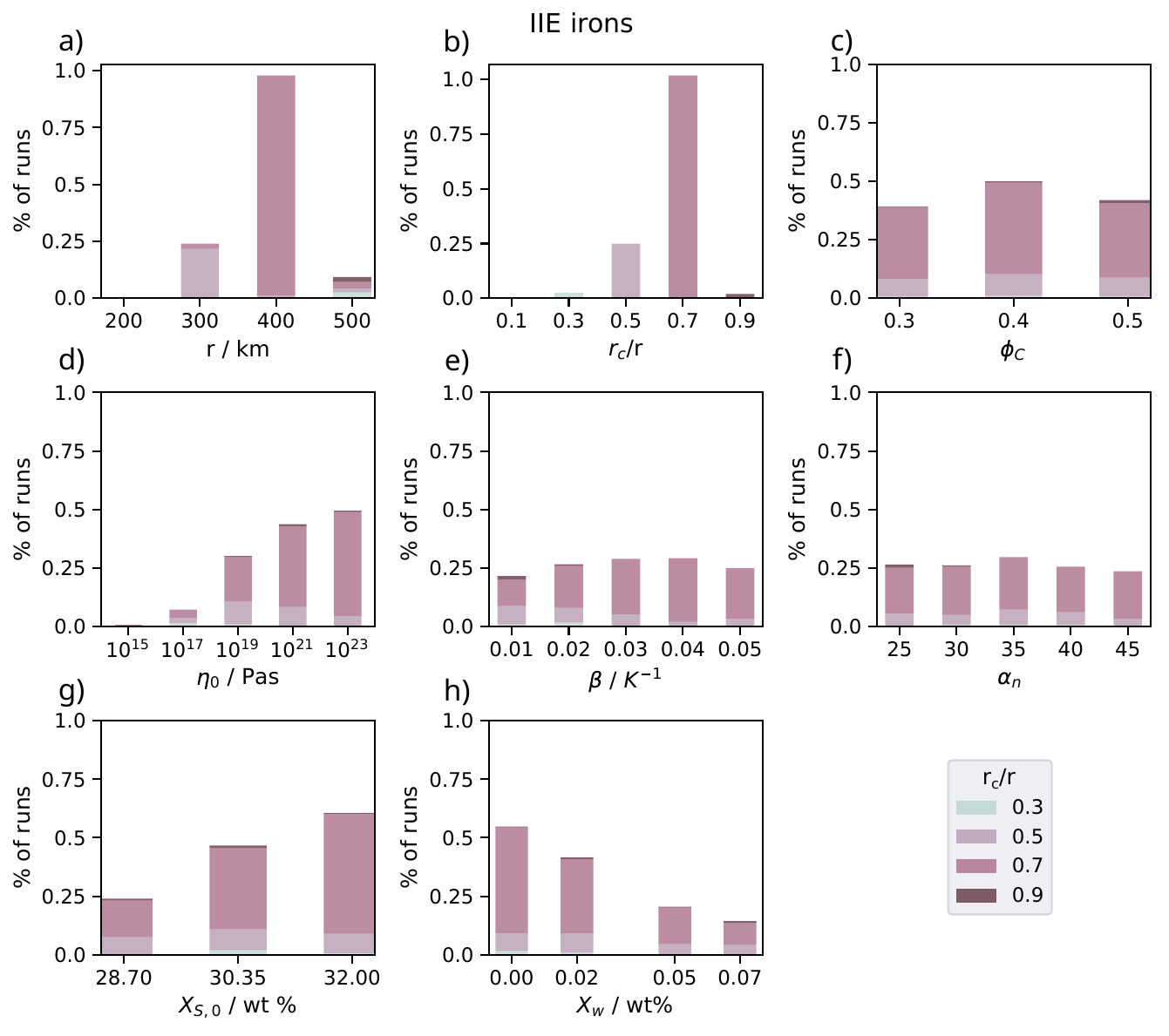}
    \caption{Distributions of input parameter values that produce thermal and dynamo histories consistent with the IIE iron meteorite record. The input parameters are planetesimal radius, $r$, core radius fraction, \rcrp, critical melt fraction, $\phi_C$, reference viscosity, $\eta_0$, Arrhenius slope, $\beta$, melt weakening exponent, $\alpha_{\mm{n}}$, initial core sulfur content, \Xsp, and mantle water content in NAMs, $X_{\mm{w}}$. Horizontal axis tickmarks indicate the possible values of the input parameters. Bars are missing at $r=200$\,km, $\frac{r_\mm{c}}{r}=0.1$ and $\eta_0=10^{15}$\,Pas, because these parameter values did not produce results consistent with the meteorite record. The y-axis is the ratio of the number runs with that parameter value that were consistent with the meteorite record compared to the total number of explored parameter combinations. Bar heights are subdivided by core radius fraction of the sucessful runs.}
    \label{fig:iie-in-hist}
\end{figure}

\begin{figure}[htbp]
    \centering
    \includegraphics[width=0.8\textwidth]{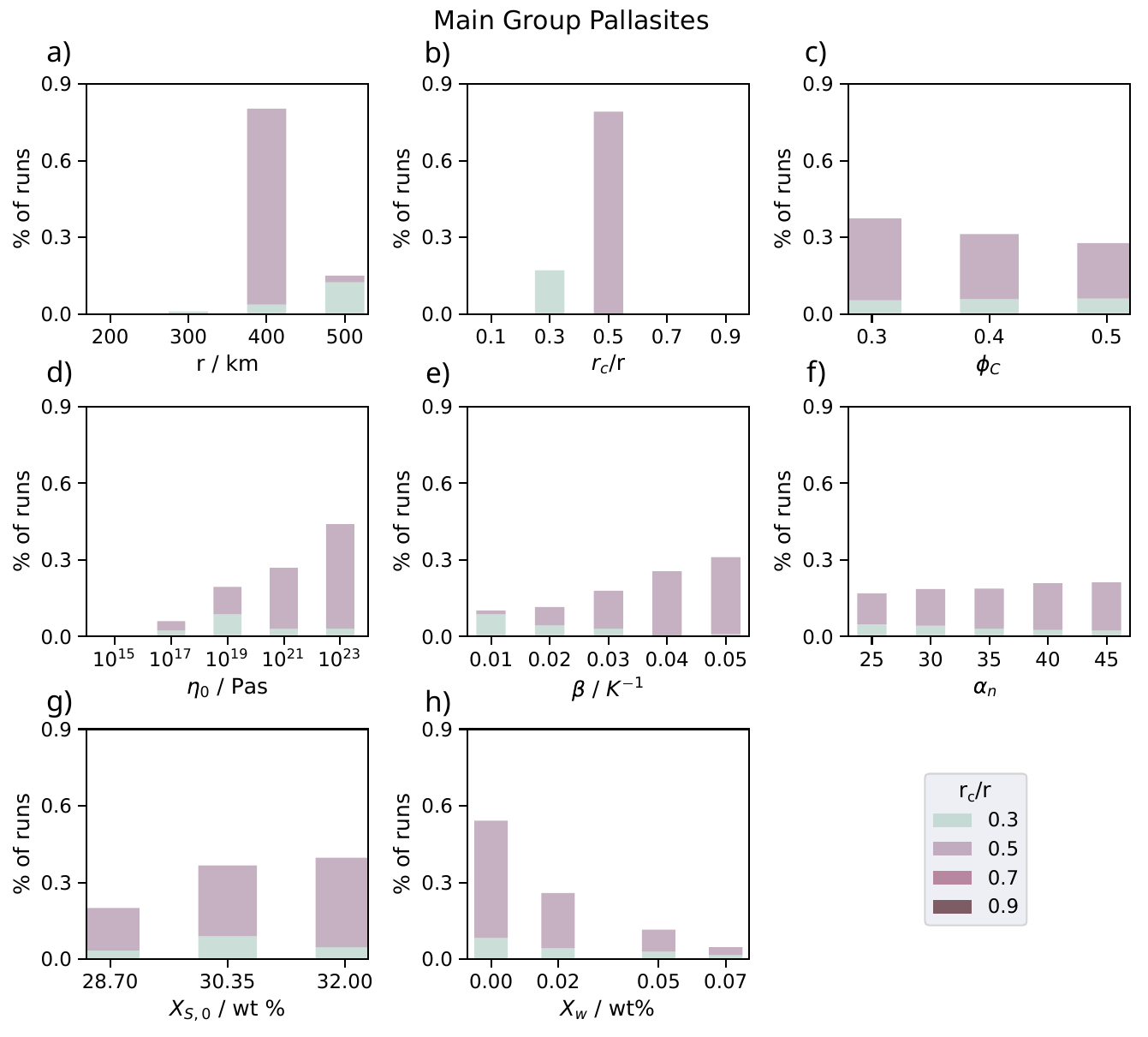}
    \caption{Distributions of input parameter values that produce thermal and dynamo histories consistent with the Main Group pallasite meteorite record. The input parameters are planetesimal radius, $r$, core radius fraction, \rcrp, critical melt fraction, $\phi_C$, reference viscosity, $\eta_0$, Arrhenius slope, $\beta$, melt weakening exponent, $\alpha_{\mm{n}}$, initial core sulfur content, \Xsp, and mantle water content in NAMs, $X_{\mm{w}}$. Horizontal axis tickmarks indicate the possible values of the input parameters. Bars are missing at $r=200$\,km, $\frac{r_\mm{c}}{r}=0.1, 0.7, 0.9$ and $\eta_0=10^{15}$\,Pas because these parameter values did not produce results consistent with the meteorite record. The y-axis is the ratio of the number runs with that parameter value that were consistent with the meteorite record compared to the total number of explored parameter combinations. Bar heights are subdivided by core radius fraction of the sucessful runs.}
    \label{fig:pal-in-hist}
\end{figure}

For both meteorite parent bodies, there is a dominant planetesimal radius (Figure \ref{fig:iie-in-hist}a and \ref{fig:pal-in-hist}a) and core radius fraction (Figure \ref{fig:iie-in-hist}b and \ref{fig:pal-in-hist}b) within the distribution of successful runs. Both bodies have the same large modal radius (400\,km), but the IIE irons have larger modal core radius fractions (0.7) than the Main Group pallasites (0.5). The Main Group pallasites are limited to smaller core radius fractions than the IIE irons because the Main Group pallasite paleomagnetic record includes a period without a dynamo. Increasing core radius fraction narrows or removes the gap in dynamo generation predicted by the models \citep{sanderson_dynamo_2026}, which is incompatible with the Main Group pallasite record. This requirement for a gap in dynamo generation also leads to fewer possible planetesimal radius and core radius fraction combinations for the Main Group pallasites (5) compared to the IIE irons (8). The core radius affects the heat flux required for a dynamo \citep{sanderson_unlocking_2025,sanderson_dynamo_2026} and together fractional core radius and planetesimal radius determine the absolute mantle thickness, which affects the mantle and core cooling rates. Therefore, changes to these parameters have a strong effect on the timing of dynamo generation. For the IIE irons, the parent body just needs a core radius and mantle thickness that results in continuous dynamo generation, while for the Main Group pallasites this combination of parameters must have the dynamo turn on within a specific time range ($\sim$170--230\,Ma after CAI formation; Table \ref{tab:out-params}) that is determined by mantle cooling rate constraints.

The reference viscosity distributions are skewed towards the highest values in both bodies, because these values provide the sufficiently slow cooling required to sustain dynamo generation for 100s of Ma. Out of the viscosity parameters, the reference viscosity, $\eta_0$, has the clearest trend (Figure \ref{fig:iie-in-hist}d and \ref{fig:pal-in-hist}d), because it has the strongest effect on the timing of dynamo generation \citep{sanderson_early_2024}. The trends in the other viscosity parameters are less pronounced (Figure \ref{fig:iie-in-hist}c, e, f and \ref{fig:pal-in-hist}c, e, f ). The Arrhenius slope, $\beta$, is shifted towards the largest values for the Main Group pallasites (Figure \ref{fig:pal-in-hist}e), because increasing the Arrhenius slope widens the gap in dynamo generation \citep{sanderson_early_2024}. A wider gap is more favourable for the Main Group pallasites, because it increases the range of times over which the null paleointensity measurements could line up with a cessation in dynamo activity.  

Both parent bodies favour the absence of water in NAMs in the mantle (Figure \ref{fig:iie-in-hist}h and Figure \ref{fig:pal-in-hist}h), but the preference is stronger in the Main Group pallasites because the models must have a gap in dynamo generation. This gap is more difficult to achieve in bodies with water in NAMs in their mantles in which core solidification begins prior to the cessation of an early thermal dynamo \citep{sanderson_dynamo_2026}. For the IIE irons, the long duration of dynamo generation favours the low initial water contents. There is no trend between mantle water content and core radius fraction. For all core radius fractions, the number of successful runs decreases with increasing mantle water content. 

\subsubsection{Initial core sulfur contents}
\begin{figure}
    \centering
    \includegraphics[width=0.5\linewidth]{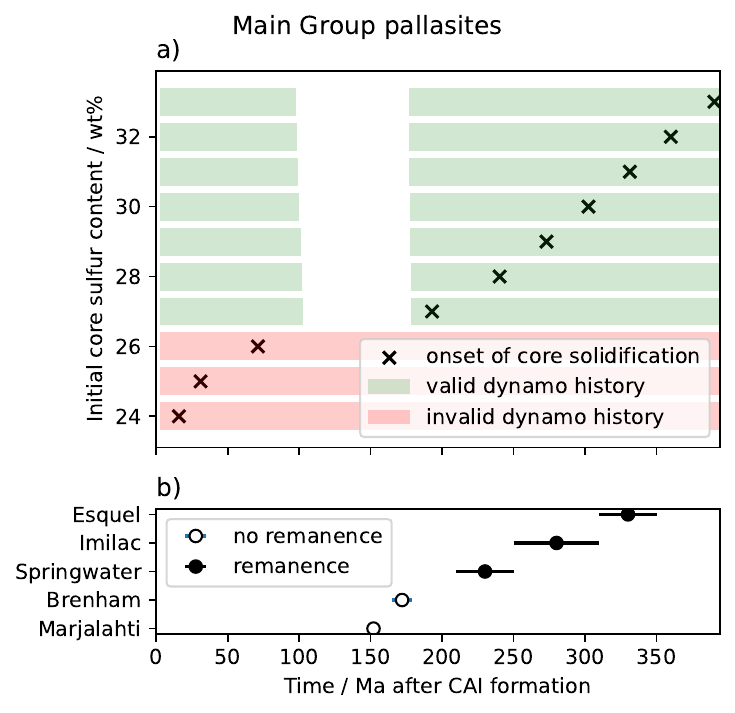}
    \caption[Dynamo duration as a function of initial core sulfur content for the modal parameters for the Main Group pallasites]{a) Dynamo duration as a function of initial core sulfur content for the modal parameters for the Main Group pallasites.  Black crosses indicate the onset of core solidification. Green bars indicate planetesimal dynamo histories that are consistent with the meteorite record, while red are those that are inconsistent. b) Mean time of remanence acquisition for each meteorite based on successful model runs (Table \ref{tab:out-params}), error bars correspond to one standard deviation. The horizontal axis limit is set by the latest time for the onset of core solidification.}
    \label{fig:pal-xs}
\end{figure}

\begin{figure}[ht]
    \centering
    \includegraphics[width=0.6\linewidth]{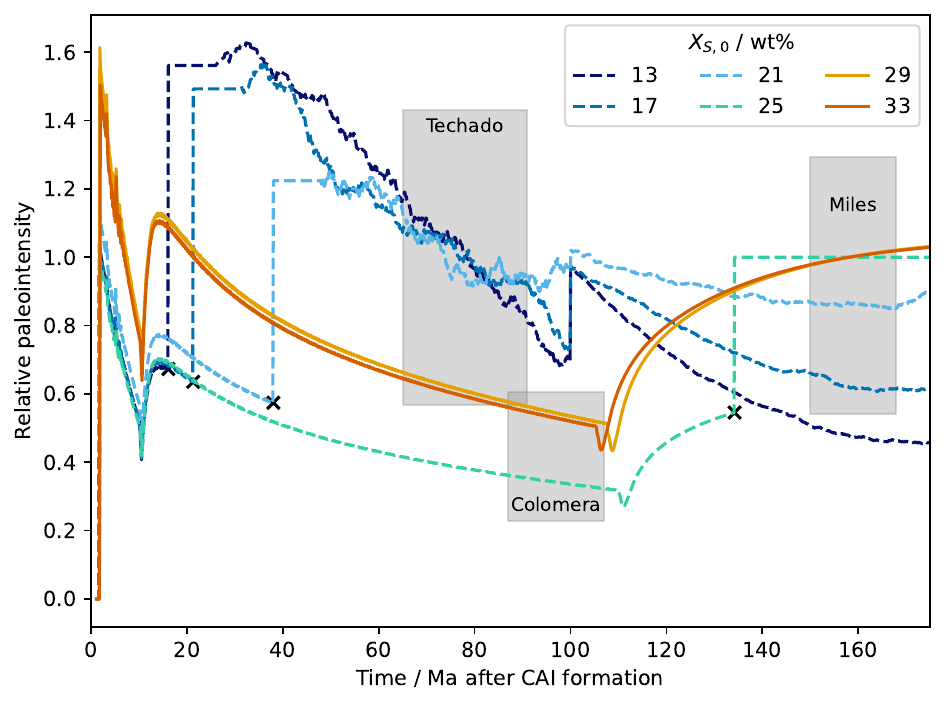}
    \caption{Model normalised paleointensity values as a function of time (lines) for a variety of initial core sulfur contents, \Xsp, compared to the normalised paleointensity data for the IIE irons (grey boxes). The height and width of the boxes indicates the uncertainty on the normalised paleointensity and time of remanence acquisition, respectively. Dashed lines indicate model runs that were incompatible with normalised paleointensity constraints, while solid lines satisfy the constraints. Black crosses indicate the onset of core solidification in each run, which is accompanied by a jump in normalised paleointensity due to the additional buoyancy flux provided by core solidification. For descriptions of other trends in the magnetic field strength see \citet{sanderson_unlocking_2025}. Planetesimal parameters are set to the IIE modal values in Table \ref{tab:params}. The x-axis limit is set by the latest time of remanence acquisition.}
    \label{fig:iie-xs}
\end{figure}

Although core solidification is no longer required to generate the remanences in the IIE irons and Main Group pallasites, the paleomagnetic data still restricts the possible values of initial core sulfur content due to its effect on the timing of core solidification and the strength of the dynamo. For the narrow range of values in Figures \ref{fig:iie-in-hist}g and \ref{fig:pal-in-hist}g, both bodies favour higher initial core sulfur contents. In a more detailed exploration, with all other planetesimal parameters held at their modal values, the initial core sulfur content for the Main Group pallasites is restricted to 27--33\,wt\%, because the Main Group pallasites require a gap in dynamo generation (Figure \ref{fig:pal-xs}). If the core has a too low initial sulfur content, it will reach the liquidus and begin solidifying before the end of the first epoch of dynamo generation. This prevents the required gap in dynamo generation because the additional buoyancy from core solidification can sustain the dynamo during the decrease in CMB heat flux around the cessation of mantle convection \citep{sanderson_unlocking_2025}. For the IIE irons, the initial core sulfur content is restricted by the jump in normalised paleointensity at the onset of core solidification (black crosses in Figure \ref{fig:iie-xs}) due to the additional buoyancy flux provided by core solidification. This jump means that only sulfur contents for which core solidification begins after all IIE irons record their remanences ($\geq$29\,wt\%) are compatible with the paleomagnetic record. 

These thresholds for initial core sulfur content will change depending on other planetesimal parameters, so are not definitive. The trends in dynamo activity with initial core sulfur content and the onset of core solidification outlined here are for modal values of all other planetesimal parameters. As a result, these trends may be valid for many parameter combinations but not all because other planetesimal parameters, such as mantle viscosity, will affect the buoyancy flux available to drive a dynamo and may prevent dynamo generation despite the onset of core solidification. This is the case for a small minority of successful model runs for the Main Group pallasites (1\%; Table \ref{tab:out-params}), where core solidification can begin before Marjalahti and Brenham cool through the temperature of remanence acquisition. For the IIE irons, core solidification begins in 10\% and 46\% of runs before Colomera and Miles record their remanences, respectively (Table \ref{tab:out-params}).

\clearpage

\subsection{Core solidification and the timing and depth of remanence acquisition}
\begin{figure}[th]
    \centering
    \includegraphics[width=1\linewidth]{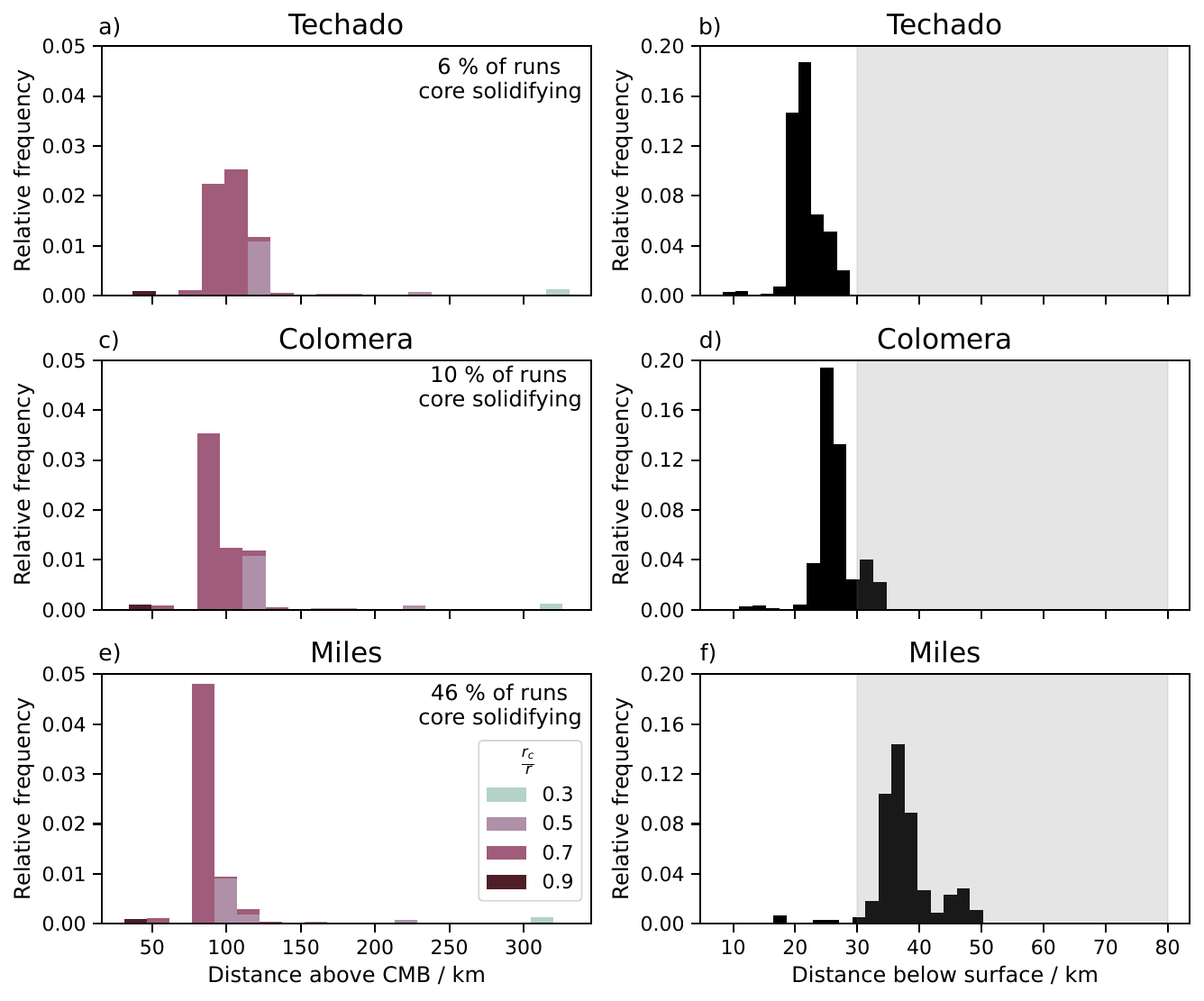}
    \caption{Distribution of meteorite formation locations within the parent body for the IIE irons displayed as distance from the CMB (left) and distance from the surface (right). The average of the upper and lower bounds on depth for each model run were used to construct the distribution. Each row is for a different meteorite. The y-axis displays the relative frequency. The mean and standard deviation for each depth and time distribution is given in Table \ref{tab:out-params}. The gaps in the middle of the distribution of distances from the CMB are arbitrary and are due to the discretisation of planetesimal radius and core radius fraction. Grey boxes indicate the depth range predicted by \citet{maurel_long-lived_2021}, which was given for all three meteorites together rather than each individual meteorite. The percentage of successful runs in which the core is solidifying is given in the left column.}
    \label{fig:iie-out-hist}
\end{figure}

\begin{figure}[t!]
    \centering
    \includegraphics[width=1\linewidth]{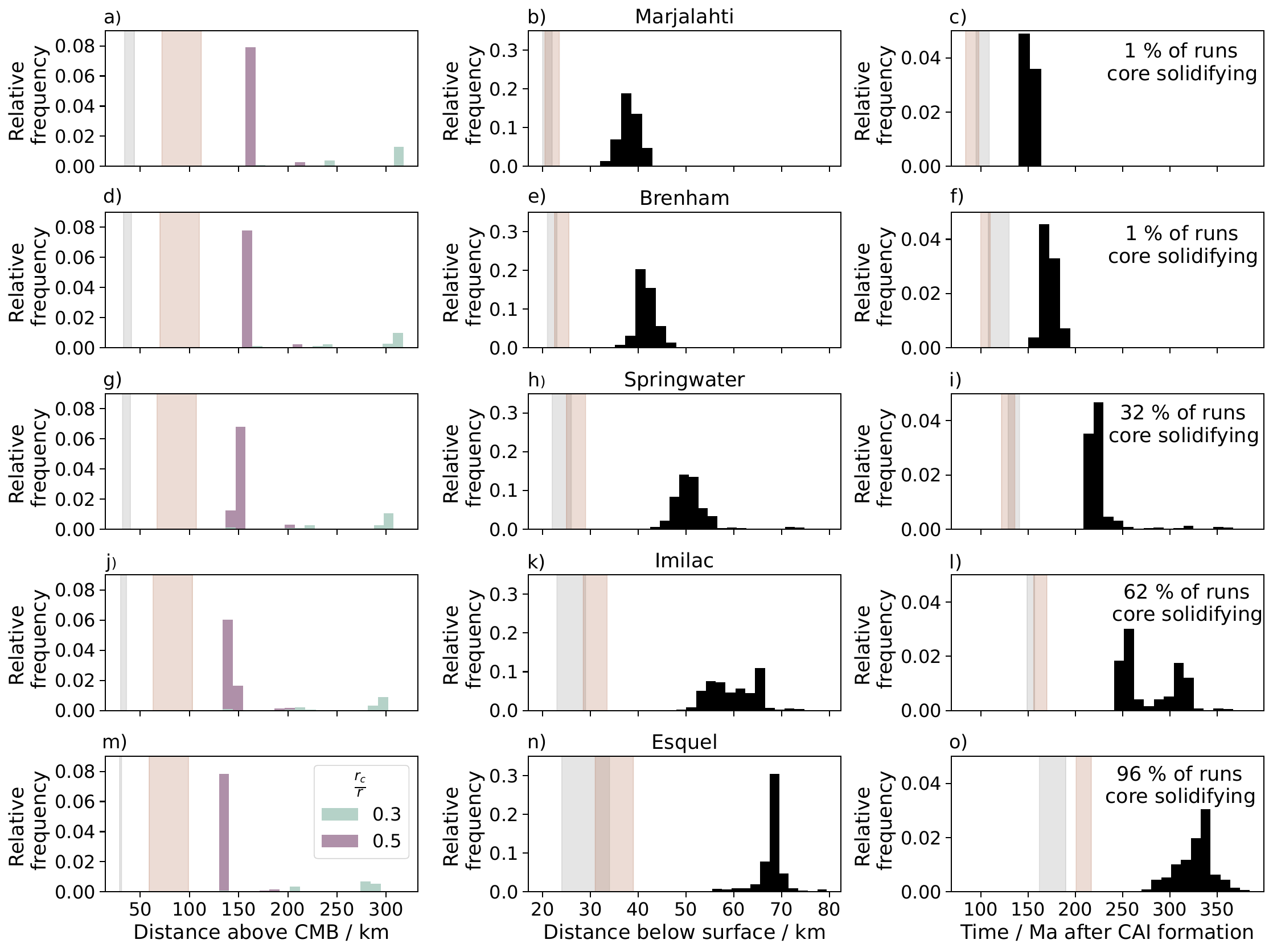}
    \caption{Distribution of meteorite formation locations within the parent body (left two columns) and time of remanence acquisition (right) for the Main Group pallasites. Origin locations are displayed as distance from the CMB (left) and distance from the surface (middle). The average of the upper and lower bounds on depth and time for each model run were used to construct the distribution. Each row is for a different meteorite. The y-axis displays the relative frequency. The boxes indicate the corresponding values for meteorite depth and time of remanence acquisition for models using inward (grey) and outward (brown) core solidification from \citet{nichols_time-resolved_2021}. The mean and standard deviation for each depth and time distribution is given in Table \ref{tab:out-params}. The percentage of successful runs in which the core is solidifying is given in the top right corner.}
    \label{fig:pal-out-hist}
\end{figure}

Our models indicate that core solidification is no longer required to generate the paleomagnetic remanences in the IIE irons and Main Group pallasites (Figures \ref{fig:iie-out-hist} and \ref{fig:pal-out-hist}). For the IIE irons, only Miles recorded a remanence during core solidification in more than 10\% of model runs. For the Main Group pallasites, the percentage of model runs for which a remanence is generated by a solidifying core increases for younger meteorites. This percentage of model runs increases from 1\% for Marjalahti and Brenham to 96\% for Esquel, so it is very likely core solidification began sometime between these groups of meteorites recording their remanences (i.e. between $\sim$170--270\,Ma after CAI formation). The standard deviation on the model ages of remanence acquisition for the Main Group pallasites (Table \ref{tab:out-params}) is the same order of magnitude as for the $\rm^{40}Ar/^{39}Ar$ ages of remanence acquisition in the IIE irons.

When meteorite cooling rates and paleomagnetic remanences are recorded, the surrounding mantle is transferring heat by conduction due to the low, subsolidus mantle temperatures at which these processes occur. As a result, the distribution of distances of a meteorite from the surface is narrow (Figure \ref{fig:iie-out-hist}b, d, f and \ref{fig:pal-out-hist}b, e, h, k, n), because the cooling rate is controlled by the conductive lengthscale (distance from the surface). In contrast, there is large variability in the distance above the CMB (Figure \ref{fig:iie-out-hist}a, c, e and \ref{fig:pal-out-hist}a, d, g, j, m), because the rate of heat transfer into the conductive lid from the convective interior is independent of thickness of the convecting mantle after $\rm ^{26}Al$ has decayed \citep{sanderson_unlocking_2025}. Therefore, the distance from a meteorite to the CMB has little effect on its measured cooling rate.
The IIE irons have a wider range of possible distances from the CMB ($\sim50$--300\,km) compared to the Main Group pallasites ($\sim$130--300\,km) because they have a wider range of  possible parent body radii and core radius fractions (Figure \ref{fig:iie-in-hist} and \ref{fig:pal-out-hist}). The dominant distances from the CMB in each distribution are for the dominant radius (400\,km) and core radius fraction ($\frac{r_\mm{c}}{r}=0.5$ or $\frac{r_\mm{c}}{r}=0.7$). Gaps in the distribution of possible distances from the CMB are due to the spacing of planetesimal radius and core fractional radius values input into the model.
\begin{table}[ht]
    \centering
    \begin{tabular}{|c|c|c|c|c|}\hline
        Meteorite & \makecell{Distance \\ from the \\ surface /km} & \makecell{Distance\\ from the \\ CMB / km} & \makecell{Time of remanence \\acquisition / Ma after \\CAI formation} & \makecell{Fraction of \\ sucessful runs\\ with remanence \\ recorded \\ during core \\solidification} \\\hline
        \multicolumn{5}{|c|}{\textbf{IIE irons}}\\\hline
        Techado & 22 $\pm$ 3 & 110 $\pm$ 40 & see Table \ref{tab:comb-con} & 0.06\\
        Colomera & 26 $\pm$ 3 & 100 $\pm$ 40 &  see Table \ref{tab:comb-con} &0.10 \\
        Miles & 37 $\pm$ 5 & 100 $\pm$ 40 & see Table \ref{tab:comb-con} & 0.46 \\\hline
        \multicolumn{5}{|c|}{\textbf{Main Group pallasites}}\\\hline
         Marjalahti &  38 $\pm$ 2 & 180 $\pm$ 50 & 152 $\pm$ 4 & 0.01\\
        Brenham & 42 $\pm$ 2 & 180 $\pm$ 50 & 172 $\pm$ 7 & 0.01\\
Springwater & 51 $\pm$ 4 & 170 $\pm$ 50 & 230 $\pm$ 20 & 0.32 \\
Imilac & 60 $\pm$ 5 & 160 $\pm$ 50 &  280 $\pm$ 30 & 0.62 \\
Esquel & 68 $\pm$ 3 & 160 $\pm$ 50 & 330 $\pm$ 20 & 0.96 \\\hline
    \end{tabular}
    \caption{Mean properties of each Main Group pallasite and IIE iron meteorite predicted by the parameter space exploration (Figure \ref{fig:iie-out-hist} and \ref{fig:pal-out-hist}). Errors are one standard deviation on the mean. The time of remanence acquisition in the IIE irons has been radiometrically dated and is a constraint on the models not a model output.}
    \label{tab:out-params}
\end{table}

\clearpage
\section{Discussion}\label{dis}
\subsection{Constraints on planetesimal properties}\label{dis:paleo-con} 
We can constrain the interior structures of the Main Group pallasite and IIE iron parent bodies using a combination of paleomagnetism and meteorite cooling rates. Most notably we find that these parent bodies had a radius of $\sim$400\,km with 50--70\% core radius fractions and dry mantles. The reference viscosity was likely high ($\geq10^{21}$\,Pas), but other viscosity parameters are poorly constrained. This is consistent with the weak effect of all viscosity parameters apart from reference viscosity on dynamo generation \citep{sanderson_early_2024}.

The depth below the surface (Figures \ref{fig:iie-out-hist}b, d, f and \ref{fig:pal-out-hist}b, e, h, k, n) and the timing of remanence acquisition (Figure \ref{fig:pal-out-hist}c, f, l, i, o) for each meteorite are tightly constrained. The standard deviation on these values is small (2--13\%, Tables \ref{tab:out-params}) despite the variation in input parameters (Figures \ref{fig:iie-in-hist} and \ref{fig:pal-in-hist}). The distance of each meteorite from the CMB (Figures \ref{fig:iie-out-hist}a, c, e and \ref{fig:pal-out-hist}a, d, g, j, m) is only constrained to within an order of magnitude, because this distance does not affect meteorite cooling rates. However, an order of magnitude is still sufficient to determine that meteorites originate far closer to the surface than the CMB in both parent bodies.

The successful parameter combinations for the Main Group pallasites and IIE irons are similarly restricted ($\sim1$\% of parameter combinations can explain paleomagnetic and cooling rate histories). For both bodies the paleomagnetic constraints are more restrictive than the cooling rate constraints (Figure S3). For the IIE irons, the normalised paleointensity is the most restrictive criterion. Of the model runs that have an active dynamo when each meteorite acquired its remanence, only 10\% have the required normalised paleointensities, whereas 90\% have the required cooling rates. For the Main Group pallasites, only 14\% of runs have the required gap in dynamo generation whereas 57\% of these runs with a gap have the correct cooling rates. Therefore, obtaining well-constrained paleointensities for the Main Group pallasites and searching for a gap in dynamo generation in the IIE iron paleomagnetic record would help the most to further narrow constraints on their parent bodies (see Section \ref{dis:future}).

\subsection{Comparison with previous studies and formation mechanisms}\label{dis:prev}
The main difference compared to previous studies is that compositional convection from core solidification is no longer required to generate the paleomagnetic remanences carried by the Main Group pallasites and IIE irons. We have quantified the fraction of model runs for which compositional convection generates the observed remanences: for the IIE irons, 6\%, 10\%, and 46\% of successful model runs predict that compositional convection contributes to the dynamo during remanence acquisition for Techado, Colomera, and Miles, respectively; for the Main Group pallasites, core solidification begins before remanence acquisition in 32\% of model runs for Springwater, 62\% for Imilac, and 96\% for Esquel. For both parent bodies, the range of possible radii and core radius fractions has changed compared to previous studies (Sections \ref{dis:diff-iie} and \ref{dis:diff-pal}). These differences arise from removing the assumption of core solidification, modelling both thermal and compositional dynamo generation, and requiring a certain vigour of convection (a supercritical magnetic Reynolds number) for dynamo generation. 

\subsubsection{IIE irons}\label{dis:diff-iie}
Two studies previously explored properties of the IIE iron parent body using a thermal model with two-stage accretion, mantle conduction, and mantle convection \citep{maurel_meteorite_2020,maurel_long-lived_2021}. These models used the timing of core solidification as a proxy for compositional convection and dynamo generation. For a model run to be successful, core solidification had to be ongoing at the $\rm ^{40}Ar/^{39}Ar$ ages of each meteorite. Therefore, the minimum core radius fraction and minimum planetesimal radius were set by the size above which a core would be partially molten at these times, rather than whether these cores were able to generate a dynamo. As a result, their minimum values of core radius fraction (13\% for 450\,km radius, 19\% for 250\,km radius) and planetesimal radius (220\,km) were lower than reported here because we show that these small planetesimals and cores are partially molten but unable to generate a dynamo at the required time. 

Based on the mixture of chondritic and achondritic silicate inclusions within the IIE irons, these meteorites are thought to originate from a partially differentiated parent body \citep{mccoy_silicate-bearing_1995,ruzicka_silicate-bearing_2014,maurel_meteorite_2020}. Therefore, \citet{maurel_long-lived_2021} imposed an upper limit on core radius fraction of 0.43, based on the metal content of the H and HH chondrites, which are isotopically similar to the chondritic inclusions in the IIE irons \citep{mcdermott_oxygen_2016}. However, our model returns a modal core radius fraction of 0.7 and 98\% of the parameter combinations consistent with the meteorite paleomagnetic record suggest a core radius fraction exceeding 0.43. A higher core radius fraction of 0.5 (19\% of successful runs) consistent with the model predictions could be achieved if the parent body's initial composition was 4\,vol.\% more metal-rich than H chondrites and the body almost completely differentiated. Depending on initial and final accretion time, it is possible for 80--90\% of a partially differentiated planetesimal's radius to be fully differentiated \citep{dodds_thermal_2021}. To achieve core radius fractions larger than 0.5 requires a fraction of the mantle to have been removed by hit-and-run collisions \citep{asphaug_similar-sized_2010,asphaug_mercury_2014}. This mantle-stripping is still consistent with an initial partially differentiated structure for the IIE iron parent body, suggested by \citet{maurel_meteorite_2020}, for two reasons. Firstly, a fraction of a planetesimal's crust can be retained during hit-and-run collisions \citep{carter_collisional_2018}, so a mantle-stripped body could have retained some chondritic material that was incorporated into the IIE irons by a later impact. Secondly, the release of hydrostatic pressure during these collisions can mix different planetary layers \citep{asphaug_similar-sized_2010}, so the IIE irons could have been formed in the same collision that stripped most of the parent body's mantle. Since 79\% of successful runs have core radius fractions greater than 0.5, we argue that IIE iron parent body likely experienced mantle stripping impacts.

\subsubsection{Main Group pallasites}\label{dis:diff-pal}
\citet{nichols_time-resolved_2021} performed the most recent and broadest exploration of possible parent body properties for the Main Group pallasites and was the first study to include paleomagnetic constraints from all five measured meteorites. \citet{nichols_time-resolved_2021} used a thermal evolution and dynamo generation model that included mantle conduction and dynamo generation by core solidification. For outward core solidification, compositional convection in the core was modelled explicitly, while for inward solidification the duration of core solidification was taken as a proxy for the period of possible dynamo generation. Dynamo generation by thermal or thermo-compositional convection was not included in the model.

Our upper limit on planetesimal radius is larger than \citet{nichols_time-resolved_2021} ($\leq$ 500\,km compared to $\leq$360\,km, respectively). In \citet{nichols_time-resolved_2021}, the upper limit was determined by the maximum radius for which core solidification would begin before Springwater acquired its paleomagnetic remanence. In our model, Springwater's paleomagnetic remanance could record thermal dynamo generation, which allows parent body radii as large as 500\,km. Similar to the IIE irons, our lower limit on planetesimal size is larger than previous studies \citep[300\,km rather than 200\,km;][]{tarduno_evidence_2012,bryson_long-lived_2015,nichols_time-resolved_2021}, because bodies with radii $<300$\,km are too small to generate a dynamo when the Main Group pallasites acquired their remanence \citep{sanderson_early_2024}. 

We demonstrate the Main Group pallasite parent body was likely large (400\,km radius) with a core radius fraction of 0.5 (Figure \ref{fig:pal-in-hist}a and b). In contrast, \citet{nichols_time-resolved_2021} suggested the parent body had a core radius fraction of 0.54--0.6 to be consistent with the cloudy zone paleointensities, or $>0.74$ to also be consistent with the olivine single crystal paleointensities. The core radius fractions consistent with cloudy zone paleointensities could still be compatible with our results, which do not resolve core radius fractions between 0.5 and 0.7. However, the core radius fractions consistent with the olivine single crystal paleointensities exceed the upper limit on core radius fraction that produces planetesimal histories consistent with the paleomagnetic record in this paper. These thin-mantled bodies have cooling rates that are too fast and have gaps in dynamo generation that are too wide for the time predicted by the model between Brenham and Springwater acquiring their remanences. 

The increase in absolute mantle thickness (larger parent body radii and smaller core radius fractions compared to previous studies) predicted here increases the depths of the Main Group pallasites and their time of remanence acquisition (Figure \ref{fig:pal-out-hist}). These larger mantles cool more slowly, so, for a given temperature, the same cooling rates are reached at greater mantle depths and later times.

The lower limit on initial core sulfur contents is higher in this study compared to \citet{nichols_time-resolved_2021}, because our model includes the initial epoch of thermal dynamo generation and convective cooling. If the initial core sulfur content is too low, core solidification begins before the cessation of the first epoch of dynamo generation and prevents the gap in dynamo generation required to explain the Main Group pallasite paleomagnetic record.

There are two dominant hypotheses for the formation of the Main Group pallasites: ferromagmatism \citep{johnson_ferrovolcanism_2020}, and injection of one planetesimal core into another's mantle during an impact \citep[e.g.,][]{yang_main-group_2010,tarduno_evidence_2012,walte_two-stage_2020,kruijer_tungsten_2022}. Ferromagmatism occurs during core solidification; pockets of sulfur-rich melt within the solidified portion of the core become pressurised and erupt into the overlying mantle up to the planetesimal's surface \citep{johnson_ferrovolcanism_2020}. There are two reasons this is an unlikely formation mechanism for the Main Group pallasites based on the results in this study. Firstly, only 1\% of successful model runs predicted that core solidification had begun by the time Marjalahti and Brenham cooled through their ordering temperature, which means these pallasites very likely formed prior to core solidification and possible ferromagmatism (Table \ref{tab:out-params}). Secondly, pallasites formed $\gtrapprox130$\,km from the core mantle boundary. This exceeds the maximum height of a ferromagmatic intrusion on a planetesimal with modal properties of the Main Group pallasite parent body by over a factor of two (see Section S3). 

Instead, the paleomagnetic record supports an impact origin for the Main Group pallasites, based on the proximity of the meteorites to the surface ($\leq$80\,km depth). This aligns with multiple other strands of evidence that support an impact origin for the Main Group pallasites: isotope disequilibrium between metal and olivine \citep{windmill_isotopic_2022,bennett_iron_2022};
comparison between high strain rate deformation experiments and textures within the meteorites \citep{walte_two-stage_2020,walte_olivine_2022,walte_mantle_2023}; similarities in $\rm^{182}W$ and Mo isotopes between Main Group pallasites and IIIAB irons suggesting metal was injected from the IIIAB iron parent body \citep{kruijer_tungsten_2022}; and agreement between thermal modelling of small impact intrusions into the mantle and the observed range of cooling rates \citep{murphy_quinlan_reconciling_2023}.  

Previously, it has been suggested that the Main Group pallasite parent body could be similar to asteroid (16) Psyche \citep{elkins-tanton_observations_2020,johnson_ferrovolcanism_2020,walte_two-stage_2020,nichols_time-resolved_2021}. For initial core sulfur contents consistent with the Main Group pallasite paleomagnetic record ($>26$\,wt\%), Psyche's core radius fraction would have to be $\geq0.9$ to be consistent with Psyche's bulk density \citep{courville_ferromagmatic_2025}. This is much larger than the upper limit on core radius fraction for the Main Group pallasites suggesting the interior structure of the Main Group pallasite parent body is likely different to asteroid (16) Psyche.

\subsection{Water content and core size of differentiated inner Solar System planetesimals}\label{dis:xw}
Both the Main Group pallasites and IIE irons are isotopically non-carbonaceous (NC), meaning they originated in the inner Solar System \citep{warren_stable-isotopic_2011,rubin_carbonaceous_2018}. Therefore, their mantle water contents and fractional core sizes could provide insight into redox conditions in the inner Solar System and how much water they retained during differentiation \citep{sanderson_dynamo_2026}. Here we build on the work of \citet{sanderson_dynamo_2026} by allowing variation in all planetesimal parameters, not just core radius fraction and mantle water content, and comparing model results to time-resolved paleomagnetic histories in detail. The distribution of successful runs is skewed to bodies with no to negligible water in their mantles: 56\% and 42\% of successful runs for the Main Group pallasites and IIE irons, respectively (Figures \ref{fig:iie-in-hist}h and \ref{fig:pal-in-hist}h). The modal core radius of the Main Group pallasite parent body (0.5) is the same as the mean core radius fraction of NC iron meteorite parent bodies \citep{grewal_accretion_2024,spitzer_comparison_2025,sanderson_dynamo_2026}.  This core radius fraction suggests the Main Group pallasite parent body likely accreted some water-ice, which raised its oxidation state relative to nebula gas with a solar C/O abundance \citep{grewal_accretion_2024,sanderson_dynamo_2026}. For the IIE iron parent body, the modal fractional core radius is larger (0.7) and was likely influenced by mantle stripping collisions so is a less reliable indicator of redox conditions during differentiation (Section \ref{dis:diff-iie}). The core radius fraction of 0.5 in the Main Group pallasites and the preference for dry mantles in both parent bodies suggests that any water in these planetesimals was degassed very efficiently during differentiation in agreement with previous studies \citep{newcombe_degassing_2023,grewal_implications_2025,peterson_reconstruction_2025,sanderson_dynamo_2026}. 

\subsection{Limitations}\label{dis:lim}
Our thermal evolution and dynamo generation model does not include impacts, gradual accretion, or partial differentiation. Impacts likely played significant roles in the formation of the IIE irons \citep[Section \ref{dis:diff-iie};][]{ruzicka_silicate-bearing_2014,maurel_meteorite_2020} and Main Group pallasites \citep[Section \ref{dis:diff-pal};][]{walte_two-stage_2020,kruijer_tungsten_2022,murphy_quinlan_reconciling_2023} and could have affected the post-impact thermal evolution by reducing the thickness of the mantle or depositing regolith on the surface \citep[e.g.,][]{bryson_long-lived_2015,walte_two-stage_2020,cambioni_formation_2026}. Recent modelling of metal-olivine intrusions within a silicate mantle suggests the range of cooling rates in the Main Group pallasites can be explained by impacts intruding small amounts of material locally without disrupting the mantle's overall thermal structure \citep{murphy_quinlan_reconciling_2023}. Combined with our prediction of a core radius fraction of 0.5 for the Main Group pallasite parent body, which does not require a mantle-stripping impact, it is not necessary to include the thermal and structural effects of impacts when modelling the entire mantle's evolution for this body. 

In contrast, the large core radius fraction that we predict for the IIE iron parent body (0.7) suggests impacts significantly affected the mantle structure of this body. Mantle stripping impacts will change the plantesimal's interior structure, accelerating core cooling and enhancing dynamo generation but are unlikely to add significant heat to the body \citep{cambioni_formation_2026}. Most planetesimals with radii $\geq400$\,km (the predicted final radius of the IIE iron parent body) continuously generate a dynamo before and during the radiometrically dated time of IIE iron remanence acquisition \citep[Figure 5 in][]{sanderson_early_2024}. Therefore, an impact prior to IIE iron remanence acquisition in $\geq400$\,km radius bodies may strengthen an already active dynamo, but will not cause dynamo onset unlike in smaller bodies like asteroid (16) Psyche \citep{cambioni_formation_2026}. As long as the impact occurred prior to all three meteorites acquiring their remanences, there will also not be a change in the normalised paleointensities predicted by the model for these three meteorites. Immediately after the mantle-stripping impact, the planetesimal mantle will be hotter than expected for its viscosity parameters and new thickness, since thinner mantles cool faster than thicker ones. However, more viscous mantles also cool more slowly than less viscous ones, so there is some degeneracy in thermal history between viscous mantles and thick mantles. This could lead our model to overestimate mantle reference viscosity, in order to keep a thin mantle hotter for longer because we do not consider the scenario when the mantle is thicker initially. However, meteorite cooling rates provide a weaker constraint on parent body properties than the two paleomagnetic criteria (Section \ref{dis:paleo-con}), which will be unaffected by impacts, so overall neglecting impacts should have little affect on the predicted size and internal structure of the IIE iron parent body.   

Gradual accretion and partial differentiation likely also played a role in the evolution of the IIE irons parent body due to the presence of achondritic and chondritic silicate inclusions within these meteorites \citep{maurel_meteorite_2020}. These processes also have been suggested to occur in the Main Group pallasite parent body \citep{walte_mantle_2023}. As discussed in \citet{sanderson_unlocking_2025}, neglecting porous, undifferentiated material close to the surface could slightly overestimate surface cooling rates, but should not change the large-scale dynamo behaviour. Partial differentiation can lead to smaller cores, which we have accounted for by exploring a range of core radius fractions.

\subsection{Proposed measurements}\label{dis:future}
\begin{figure}
    \centering
    \includegraphics[width=1\linewidth]{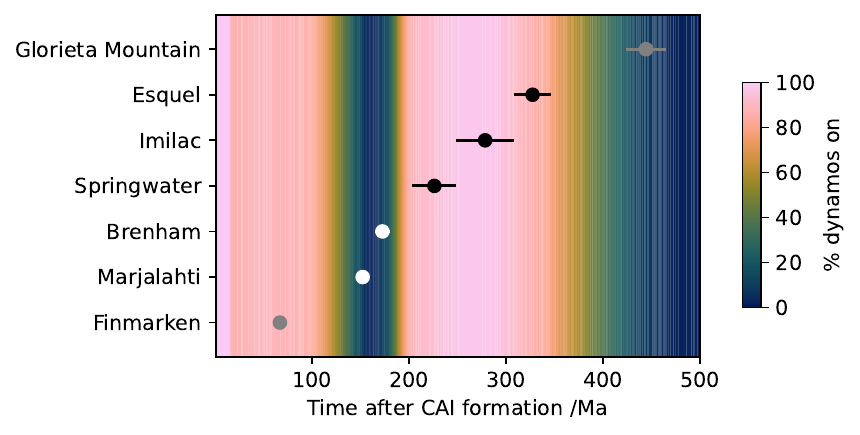}
    \caption{Predicted dynamo history for the Main Group pallasites. The colourmap indicates the percentage of successful model runs for which the dynamo is on at a given time. The white, black, and grey points correspond to the timing of remanence acquisition for each meteorite with zero, non-zero and unmeasured remanences, respectively. Errorbars on the points are one standard deviation on the mean and are less than the width of the point if not shown.}
    \label{fig:pal-predict}
\end{figure}

As discussed in Section \ref{dis:paleo-con}, reducing the uncertainty on Main Group pallasite paleomagnetic measurements so a normalised paleointensity criterion can be adopted in this analysis would be key to further improving constraints on the Main Group pallasite parent body. Reducing this uncertainty would both require additional XPEEM measurements on Marjalahti, Brenham, Imilac, and Esquel from one to at least three rotations to remove the uncertainty on the direction of the primordial field \citep{bryson_paleomagnetic_2019,nichols_time-resolved_2021} and a new model of cloudy zone remanence acquisition that accounts for magnetic interactions between cloudy zone islands when converting these measurements into paleointensities \citep{mansbach_resolving_2026}. Furthermore, paleomagnetic and cooling rate measurements of the Glorieta Mountain and Finmarken meteorites would also provide additional time points in the Main Group pallasites paleomagnetic history. Glorieta Mountain has a slower cooling rate \citep[$2.5\pm0.3$\,K/Ma at 925\,K;][]{yang_main-group_2010} and hence deeper formation depth and later remanence acquisition time than Esquel. Therefore, this meteorite could record the end of the dynamo (Figure \ref{fig:pal-predict}). Finmarken has a much faster cooling rate \citep[$18.7\pm1.2$\,K/Ma at 925\,K;][]{yang_main-group_2010}, shallower formation depth, and earlier time of remanence acquisition than Marjalahti. As a result, this meteorite could record a first epoch of dynamo generation on the parent body (Figure \ref{fig:pal-predict}) and help determine the duration of the gap in dynamo generation. These additional measurements could better constrain planetesimal viscosity parameters and mantle water content depending on the combination of zero and non-zero paleointensities recorded by these two meteorites (Figures S4--S7).

There is also very little thermochronology for the Main Group pallasites. Because there is no feldspar in the Main Group pallasites \citep{wasson_main-group_2003}, the feldspar $\rm ^{40}Ar/^{39}Ar$ system used to date remanence acquisition in the IIE irons \citep{bogard_chronology_2000} cannot be applied to the Main Group pallasites. $\rm ^{53}Mn/^{53}Cr$ dating has been performed on minor chromite phases (closure temperature $\approx$1300\,K) but only for Brenham, so could be attempted for the other meteorites that have been paleomagnetically measured \citep{windmill_isotopic_2022}. 

For the IIE irons, finding a gap in dynamo generation would provide the most informative constraints on this parent body. Unfortunately, there are currently no good candidates for additional paleomagnetic measurements. Out of the IIE irons with measured $\rm ^{40}Ar/^{39}Ar$ ages, Weekeroo Station formed at the same time as Techado, so would not provide additional information; while the ages of Watson and Netscha{\"e}vo were reset by an event $>800$\,Ma after CAI formation \citep{bogard_chronology_2000}. Therefore, the first step should be to measure $\rm^{40}Ar/^{39}Ar$ ages for more IIE irons. Once these ages are obtained, any meteorites younger than Miles to search for a gap or older than Techado to try find the cessation of dynamo generation should be the first target for paleomagnetic analysis.

\section{Conclusions}\label{conc}
Time-resolved paleomagnetic records of stony-iron and silicate-bearing iron meteorites can been used to determine the interior structure and evolution of their differentiated parent bodies and understand the formation of these meteorites. However, previous studies utilising the paleomagnetic records of the Main Group pallasites and IIE irons in this way assumed that the magnetic fields recorded by these meteorites were generated by core solidification; an assumption recently shown to be invalid \citep{sanderson_early_2024}. Here, we explored the consequences of removing this assumption for the formation histories and parent body properties of the IIE irons and Main Group pallasites. We used the planetesimal thermal evolution and dynamo generation model of \citet{sanderson_unlocking_2025} to calculate planetesimal thermal and dynamo histories over a wide range of input parameters. Then, we compared model outputs to paleomagnetic data, metallographic cooling rates, and radiometric ages to determine possible values of parent body properties, including planetesimal size, core radius, and mantle viscosity. We also estimated each meteorite's formation distance from the surface and CMB (and timing of remanence acquisition for the Main Group pallasites). This has revealed the following insights about these two parent bodies: \begin{itemize}
    \item Core solidification is no longer a requirement for producing any of the observed paleomagnetic remanences. The percentage of successful model runs which predicted that an individual meteorite recorded a remanence generated by core solidification varies between 6--46\% and 32--96\% for the IIE irons and Main Group pallasites, respectively.
    \item The Main Group pallasite and IIE iron parent bodies were likely large ($\sim$400\,km radius) with core radius fractions of 0.5 and 0.7, respectively. This suggests the IIE iron parent body experienced mantle-stripping impacts at some point in its history based on compositional constraints for initial metal fraction. 
    \item The possible combinations of planetesimal radius and core radius fraction are more limited for the Main Group pallasites than the IIE irons because of the observed period without dynamo generation in Main Group pallasite paleomagnetic record. 
    \item The Main Group pallasites formed $\leq80$\,km below the surface and $\gtrapprox$130\,km from the CMB. This suggests these meteorites formed by impacts rather than ferromagmatism.
    \item The paleomagnetic records of the Main Group pallasites and the IIE irons support efficient degassing of water during planetesimal differentiation and are consistent with their formation in the NC reservoir.
\end{itemize}
Overall, combining meteorite paleomagnetism and cooling rates with thermal evolution and dynamo generation models can provide valuable insights into the long-term evolution of differentiated planetesimals, their interior structures, and metal-silicate mixing on these bodies.

\printcredits 

\section*{Declaration of competing interest}
The authors declare that they have no known competing financial interests or personal relationships that could have appeared to influence the work reported in this paper.

\section*{Data availability}
The dynamo and thermal evolution model and the summary output files from the model runs required to recreate the results in this paper are publicly available in the \citet{sanderson_refined_2026} Github repository. 

\section*{Acknowledgements}
H.R.S. acknowledges funding on a Natural Environment Research Council studentship NE/S007474/1, an Exonian Graduate Scholarship from Exeter College, University of Oxford and funding from the Research Council of Norway through the Centres of Excellence funding scheme, project number 332523 (PHAB). JFJB acknowledges funding from the UKRI Research Frontier Guarantee program EP/Y014375/1. For the purpose of open access, the authors have applied a Creative Commons Attribution (CC BY) licence to any Author Accepted Manuscript version arising. 

\bibliographystyle{cas-model2-names}
\bibliography{references}
\end{document}